\documentclass[prb,twocolumn,preprintnumbers,amsmath,amssymb]{revtex4}

\usepackage{graphicx}
\usepackage{dcolumn}
\usepackage{bm}

\begin{document}

\title{
Electronic Excitations from a Perturbative LDA+$GdW$ Approach}
\author{Michael Rohlfing}
\email{Michael.Rohlfing@uos.de}
\affiliation{
  Fachbereich Physik, Universit\"at Osnabr\"uck, 
  Barbarastra{\ss}e 7, 49069 Osnabr\"uck, Germany}

\date{\today}

\begin{abstract}

We discuss an efficient approach to excited electronic states within 
ab-initio many-body perturbation theory (MBPT).
Quasiparticle corrections to density-functional theory result from the 
difference between metallic and non-metallic dielectric screening.
They are evaluated as a small perturbation to the DFT-LDA band 
structure, rather than fully calculating the self energy 
and evaluating its difference from the exchange-correlation potential.
The dielectric screening is desribed by a model, which applies to bulk
crystals, as well as, to systems of reduced dimension, like molecules, 
surfaces, interfaces, and more.
The approach also describes electron-hole interaction.
The resulting electronic and optical spectra are slightly less accurate
but much faster to calculate than a full MBPT calculation.
We discuss results for bulk silicon and argon, for the 
Si(111)-(2$\times$1) surface, the SiH$_4$ molecule, an argon-aluminum 
interface, and liquid argon.

\end{abstract}


\maketitle

\section{Introduction}

Many-body perturbation theory (MBPT) has become the state-of-the-art for
excited states in electronic-structure theory.\cite{Onida02,Rohlfing00}
Starting from a density-functional theory (DFT) calculation, the $GW$ 
method\cite{Hedin69} and its combination with the Bethe-Salpeter 
equation\cite{Onida02,Rohlfing00}
(BSE) allow to investigate the spectra of electrons, holes, and 
correlated electron-hole pairs.
The great success of MBPT is based on the systematic incorporation 
of Coulomb interaction and polarization effects on all length scales, 
which is not considered in most other electronic-structure approaches.
The significant computational cost of MBPT, however, still
constitutes a major obstacle for the widespread use of the method.
This holds in particular for larger-scale systems, like defects, 
hybrid systems, adsorbates, nanostructures, and others.
In this paper we propose a dramatic reduction of the computational 
requirements of MBPT.
As a result, the excellent precision of standard  $GW$ and $GW$+BSE 
calculations is slightly reduced, but instead the treatment of much
larger systems becomes possible, thus allowing the investigation
of spectroscopic features that might be inaccessible otherwise.

\begin{figure}
\scalebox{0.45}{\includegraphics{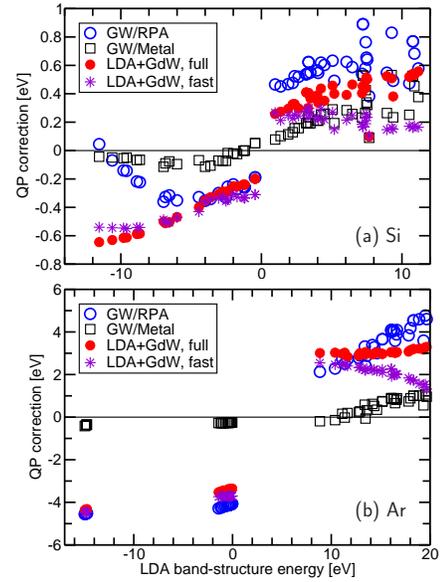}}
\caption{
\label{fig_1}
(Color online)
Quasiparticle corrections of (a) bulk Si and (b) bulk Ar.
The open circles ($\circ$) denote standard $GW$ data within RPA 
screening.
The squares ($\Box$) result from a standard
$GW$ calculations, but based on a {\em metallic} dielectric model 
function (see text).
The filled circles ($\bullet$) result from the present LDA+$GdW$
approach, employing the same basis and band-summation details as
in the standard $GW$ data.
The asterisks ($\ast$, ''fast'') result from LDA+$GdW$ with 9 plane 
waves (15 plane waves for Ar) and 8 bands, only (see text).
}
\end{figure}

As key ingredient
we exploit the observation that for many systems MBPT, when 
carried out by (wrongly) assuming {\em metallic dielectric screening}, 
approximately reproduces the band structure of the underlying DFT 
calculation (when employing the local-density approximation, LDA).
This had already been observed by Wang and Pickett,\cite{WP83}
as well as, by Gygi, Baldereschi, and Fiorentini 
\cite{Gygi89,Fiorentini95}
and was subsequently exploited for model QP calculations for various
materials.\cite{Massidda95,Continenza99}
As illustration, Fig. \ref{fig_1} shows quasiparticle (QP) corrections 
for silicon (Si) and solid argon (Ar).
The open circles ($\circ$) result from a conventional 
$GW$ calculation (with standard RPA dielectric screening,
''$GW$/RPA''), 
yielding the well-known opening of the band gap (by 0.7 eV for Si and
6.1 eV for Ar).
The squares ($\Box$), on the other hand, 
come from $GW$ calculations which employ {\em metallic} screening;
these QP shifts are close to zero (at least for states near the 
Fermi level).
The ''metallic'' dielectric screening is simulated by a dielectric 
model function.\cite{Bechstedt81,Levine82,Hybertsen88,Falter91,Cappellini93}
Here we use a model based on that of Bechstedt, Enderlein and 
Wischnewski,\cite{Bechstedt81} slightly modified (see 
Sec. \ref{subsec_model}) for broader applicability.
Such models are controlled by a few parameters, most importantly by 
the macroscopic dielectric constant, $\epsilon_\infty$.
Setting $\epsilon_\infty$$\equiv$$\infty$ turns the 
screening into that of a metal.

If the $GW$ method with metallic screening, $W_{metal}$, reproduces the 
DFT-LDA band structure, one can arrive at the true QP band structure by
adding a self energy
\begin{equation}
\Delta \Sigma := iG_1(W-W_{metal}) = i G_1 \Delta W
\label{eq_0}
\end{equation}
to the DFT-LDA Hamiltonian\cite{WP83,Gygi89,Fiorentini95}.
This procedure, which we label ''LDA+$GdW$'' throughout the paper
(see below for details), yields the filled circles ($\bullet$) in 
Fig. \ref{fig_1}.
These data agree fairly well with the $GW$/RPA data, at least in 
the important region near the fundamental gap;
as a rule of thumb, we find that the gaps from 
$\Delta \Sigma := iG_1(W-W_{metal})$
agree with experiment to about 10 \%,
which largely corrects the DFT-LDA band-gap error of 30-50 \%.
Most importantly, long-range polarization effects are included, 
which is completely missing from the short-sighted DFT-LDA.
This systematic improvement due to $\Delta\Sigma$ is much more 
important than plain agreement of band-structure data with experiment.

While Eq. (\ref{eq_0}) describes single-particle states of electrons 
and holes, a straight-forward extension to coupled electron-hole pairs 
and their optical response is easily possible by calculating the 
electron-hole interaction (from $W$) and solving the Bethe-Salpeter 
equation, BSE.\cite{Onida02,Rohlfing00}
Since the self-energy considered in this work is in principle still the 
one resulting from the $GW$ approximation (GWA), the corresponding 
approximations to the electron-hole interaction are meaningful in the 
context of LDA+$GdW$, as well.

As an important consequence of the above
findings, the calculation of $GW$-like band structures
via $\Delta\Sigma$ allows for a tremendous gain in numerical efficiency
in four respects: ({\it i}) the use of model dielectric functions,
({\it ii}) small basis-set requirement, ({\it iii}) small band-summation
requirement, and ({\it iv}) weak influence of dynamical effects.
The underlying reason for all four issues is that the current approach 
calculates QP corrections to DFT-LDA as a {\em perturbation}.
Most other $GW$ implementations simply replace $V_{xc}$ by 
$\Sigma^{GW}$, both of which are in the order of magnitude of --10 eV
or more.
In order to get their difference to within 0.1 eV, the $GW$ calculation
must be carried out with numerical precision better than 1 \%.
Our present approach, on the other hand, starts direcly at the
QP {\em correction} (i.e., $\Delta\Sigma$), which is much smaller 
in magnitude ($\sim$1 eV) and much more robust.
Here it is fully sufficient to evaluate all quantities to within 
10 \%, only, to achieve the same final numerical accuracy in the 
band structure.

The approach to be proposed in this paper is similar to the method
by Gygi, Baldereschi and Fiorentini,\cite{Gygi89,Fiorentini95} who 
employed their perturbative $GW$ method for the comprehensive analysis 
of bulk crystals.
As a key difference to their approach, here we employ a 
different, more general model dielectric 
function which is flexible enough to also describe composite systems
containing metals, non-metals, molecules, surfaces,
interfaces, and more.
As illustration, we discuss in this paper bulk materials, a 
semiconductor surface, a molecule, a metal-insulator junction, and
a disordered insulator.
In all cases our approach yields spectroscopic data of high quality
(although slightly less accurate than the corresponding full $GW$ or 
$GW$/BSE calculation), demonstrating an appealing combination of 
predictive power, broad applicability, and numerical eficiency.

The paper is organized as follows.
In Sec. \ref{sec_method} we discuss the computational approach and
the dielectric model function employed in this work.
In Sec. \ref{sec_bulk} characteristic results for bulk silicon and 
bulk argon are discussed.
Sec. \ref{sec_more} presents results for more complicated systems,
indicating the potential of the method.
The paper is concluded by a summary in Sec. \ref{sec_summary}.

\section{Theoretical approach}
\label{sec_method} 

In this section we discuss the computational method used in this work,
including its practical realization and underlying physical principles.

\subsection{Perturbative quasiparticle corrections}

Ab-initio quasiparticle (QP) band structures result from the electron 
self-energy operator $\Sigma(E)$.
The state-of-the-art approach to $\Sigma$ is given by Hedin's $GW$ 
approximation,\cite{Hedin69} which is usually evaluated and employed
on top of an underlying density-functional theory (DFT) calculation.
The typical procedure employs DFT data to generate the single-particle
Green function $G_1$ and the screened interaction $W$ (usually within 
the random-phase approximation).
Thereafter, the resulting self-energy operator $\Sigma = i G_1 W$ 
replaces the DFT exchange-correlation potential, $V_{xc}$,
arriving at a QP Hamiltonian of
\begin{equation}
\label{eq_1}
\hat{H}^{QP} := \hat{H}^{DFT}  + i G_1 W - V_{xc}
\end{equation}
This procedure is commonly labelled ''many-body 
perturbation theory'' (MBPT). 
However, this does {\em not} mean that  Eq. (\ref{eq_1}) would be
evaluated truly perturbatively in the sense that the 
smallness of the difference ($i G_1 W - V_{xc}$) would be exploited.
Instead, both terms, $i G_1 W$ and $V_{xc}$, are evaluated separately,
taking their difference afterwards.
The QP corrections are thus obtained as (small) differences between
two (rather large) quantities, both of which have to be evaluated
independently and with high precision.
Simply speaking, in order to get their difference (often $\sim$1 eV)
to within 0.1 eV (which is the accuracy expected from MBPT),
both $i G_1 W$ and $V_{xc}$ (being of the order of $\sim$ $-$10 eV 
or more) need to be evaluated with a precision of 1\%.
The underlying reason for this problem is the quite different
conceptual origin of the two terms, which makes it difficult to
formulate their difference in analytic terms.\cite{Godby88}

Fortunately, there does exist some pragmatic link between $\Sigma$ and 
$V_{xc}$:
The self energy of the homogeneous electron gas (as a function of
the energy of a given state) is nearly constant (see Ref. 
\onlinecite{Hedin69}) and thus nearly coincides with $V_{xc}$.
[Note that this might not be truly fulfilled by approximations to
$\Sigma$, like the GWA, which might suffer from offsets.]
This behavior is reflected by the observation that in bulk metals,
QP corrections (from GWA) to DFT-LDA band structures are very 
small.\cite{Nor87_89,Mahan89}
In other words, DFT (at least within the local-density approximation, 
LDA) does contain correct spectral properties of the quasiparticles,
at least for homogeneous systems.
Within LDA, however, these spectral properties are by construction 
still those of the metallic system (jellium) from which the LDA 
exchange-correlation data originate, and this metallic behavior (in 
particular, metallic screening) is a built-in property of the 
exchange-correlation potential, even when applied to non-metallic 
systems.
A generalization of this statement would imply that 
$V_{xc}\approx i G_1 W_{metal}$ (provided that $i G_1 W$ is a good 
approximation to $\Sigma$) with the appropriate $W_{metal}$ (i.e. 
metallic screening).
In fact, $GW$ studies employing metallic screening (including ours,
see Fig. \ref{fig_1}) confirm that 
$V_{xc} |n{\bf k}\rangle \approx i G_1 W_{metal}|n{\bf k}\rangle$
for most electronic states $|n{\bf k}\rangle$.

Based on the working hypothesis that for non-homogeneous, non-metallic
systems the largest difference to metallic behavior is the difference
in {\em screening}, and employing $V_{xc}\approx i G_1 W_{metal}$,
one arrives at the QP Hamiltonian
\begin{equation}
\label{eq_2}
\hat{H}^{QP} \approx \hat{H}^{DFT-LDA}  + i G_1 (W-W_{metal}) \quad ,
\end{equation}
in which $\Delta\Sigma = i G_1 (W-W_{metal})$ acts as a self energy,
yielding QP corrections (cf. Eq. (\ref{eq_0})).
The most important change to Eq. (\ref{eq_1}) is the fact that
Eq. (\ref{eq_2}) no longer evaluates the difference between the
self energies (given by $i G_1 W$ and $V_{xc}$), but the difference
in screening: ($W-W_{metal}$).
This difference is much simpler and faster to 
treat than the difference between self energies.
Note that on the other hand, the final accuracy of the QP band 
structure might be less than 0.1 eV because the entire approach,
although being much more efficient from a numerical point of view,
is based on the assumption that $V_{xc}\approx i G_1 W_{metal}$,
meaning a further approximation in addition to the $GW$ approximation.
The assumption that $V_{xc}\approx i G_1 W_{metal}$ should be checked
carefully for each system class.

As discussed below, the use of Eq. (\ref{eq_2}) allows for several
numerical simplifications (see Sec. \ref{subsec_numerics}),
leading to a higher efficiency than conventional 
$GW$ calculations, allowing to tackle more complex systems.
One of the most important facilitations is the use of model
dielectric functions (see next section) instead of calculating the 
screening within the random-phase approximation.

\subsection{Model dielectric function}
\label{subsec_model}

The calculation of the dielectric function within the random-phase 
approximation (RPA), which is the common procedure within
MBPT, is one of the bottlenecks of the method.
A simplified evaluation of the dielectric function is
an important contribution to improving the efficiency of MBPT
(even without the considerations of the previous section).
For this reason model dielectric functions are sometimes employed 
to avoid the RPA.\cite{Northrup93,Benedict98,Schmidt99,Benedict99,Hahn02,Furthmueller02}
In the present context, the use of models is also mandatory for another
reason:
The key ingredient of the present theory is the difference between
the correct screening of the (non-metallic) system and its 
(hypothetical) metallic counterpart.
This is only useful and well-defined if both types of screening
result from the same approach, allowing to tune the screening
from ''correct'' to ''metallic'' in a seamless manner.
It is, however, unclear how the RPA could be used to 
simulate metallic behavior of a non-metallic system.
An appropriate model is therefore a necessity of the current approach.

Examples are the models proposed by Bechstedt, Enderlein, Wischnewski,
and Falter and by Levine, Hybertsen and 
Louie.\cite{Bechstedt81,Levine82,Hybertsen88,Falter91,Cappellini93}
We have tested these models in the present context and find that they
yield essentially the same results as the ones to be discussed below.
In their original form, however, these models have one significant 
disadvantage which may hinder their application to more complex 
systems: they were formulated for systems that are characterized by 
one common dielectric constant without spatial variation.
This makes it difficult to employ them for systems in which the 
screening shows spatial variation, like interfaces, molecules, etc.

Instead we propose a model that is based on a combination of
localized and delocalized quantities.
The system may consist of $N$ atoms (at positions 
${\boldmath\mbox{$\tau$}}_j$) 
in a (periodically repeated) unit cell or supercell of volume $V$, 
with reciprocal lattice vectors {\bf G}. 
To each atom we attribute a static charge-density response
$\chi^{(j)}$ (see below) and an effective volume $V_j$.
The dielectric function of the whole system is then obtained as
\begin{equation}
\epsilon_{{\bf G},{\bf G}'}({\bf q}) = 
  \delta_{{\bf G},{\bf G}'} + \frac{1}
    {|{\bf q}+{\bf G}||{\bf q}+{\bf G}'|}
 \sum_{j=1}^N  \frac{V_j}{V} 
  \chi^{(j)}_{{\bf G},{\bf G}'}({\bf q})  \quad .
\label{eq_3}
\end{equation}
The volume attributed to each atom controls the weight which the
atom contributes to the response.
The transformation from the charge-density response to the dielectric
function further involves a convolution with the Coulomb interaction,
i.e. the multiplication by 
$1 /(|{\bf q}+{\bf G}||{\bf q}+{\bf G}'|)$ in Eq. (\ref{eq_3}).
Note that we work with a symmetrised dielectric 
function.\cite{Rohlfing95,Rohlfing96}

It was suggested by Bechstedt {\em et al.} to describe the 
charge-density response of a (homogeneous) system with dielectric 
constant $\epsilon_\infty$ by\cite{Bechstedt81}
\begin{equation}
f(Q;\bar{\rho},\epsilon_\infty) = \left[\frac{1}{\epsilon_\infty-1} + 
    \frac{Q^2}{q_{TF}^2(\bar{\rho})} + 
    \frac{Q^4}{\omega_P^2(\bar{\rho})} \right]^{-1}
\label{eq4}
\end{equation}
where the Thomas-Fermi wave number $q_{TF}$ and plasma frequency
$\omega_P$ depend on the average electron density $\bar{\rho}$.
Eq. (\ref{eq4}) is related to the Lindhard dielectric function.
In combination with Eq. (\ref{eq_3}) (for homogeneous systems,
disregarding the summation over atoms), $\chi$=$Q^2\cdot f$ 
would describe the dielectric function.
In particular, for $Q$$\to$0 one would correctly obtain
$\epsilon(Q)$$\to$$\epsilon_\infty$ 
(if $\epsilon_\infty$$<$$\infty$, i.e. for non-metals)
or
$\epsilon(Q) \to 1 + q_{TF}^2/Q^2$ 
(if $\epsilon_\infty$=$\infty$, i.e. for metals), respectively.
The generalization to {\em non}-homogeneous systems is less clear.
While the large-$Q$ behavior,
which reflects the short-range reaction
of electronic charge to external fields on the sub-atomic length scale,
appears realistic for non-homogeneous systems as well,
the realistic incorporation of atomic-length-scale charge-density 
variation and of local fields (i.e. off-diagonal matrix elements of 
$\chi^{(j)}_{{\bf G},{\bf G}'}({\bf q})$) is less clear.

Here we propose to model the charge-density response attributed to 
each atom by:
\begin{eqnarray}
\lefteqn{\mbox{atom with metallic response:}} & & \nonumber\\
  \chi^{(j)}_{{\bf G},{\bf G}'}({\bf q}) & = &
    \sqrt{f(|{\bf q}+{\bf G}|;\bar{\rho}_j,\infty)f(|{\bf q}+{\bf G}'|;\bar{\rho}_j,\infty)} \cdot 
\nonumber\\
   & & \hspace*{-1cm} \cdot |{\bf q}+{\bf G}||{\bf q}+{\bf G}'|
      \cdot e^{-\gamma_j({\bf G}'-{\bf G})^2}
            e^{i({\bf G}'-{\bf G}){{\boldmath\mbox{$\tau$}}}_j}
\label{eq5}
\\
\lefteqn{\mbox{atom with nonmetallic response:}} & & \nonumber\\
  \chi^{(j)}_{{\bf G},{\bf G}'}({\bf q}) & = &
    \sqrt{f(|{\bf q}+{\bf G}|;\bar{\rho}_j,\epsilon_j)f(|{\bf q}+{\bf G}'|;\bar{\rho}_j,\epsilon_j)} \cdot 
\nonumber\\
   & & \hspace*{-1cm} \cdot ({\bf q}+{\bf G})({\bf q}+{\bf G}')
      \cdot e^{-\gamma_j({\bf G}'-{\bf G})^2}
            e^{i({\bf G}'-{\bf G}){{\boldmath\mbox{$\tau$}}}_j}
\label{eq6}
\end{eqnarray}
In both cases, the factor 
$\sqrt{f(|{\bf q}+{\bf G}|)f(|{\bf q}+{\bf G}'|)}$ is a reasonable
average of the large-$Q$ behavior in directions
$({\bf q}+{\bf G})$ and $({\bf q}+{\bf G}')$.
The phase factor for each atom results from the position of the
atom within the unit cell or supercell.
The factor $\exp[-\gamma_j({\bf G}'-{\bf G})^2]$ describes the
spatial extent of the charge density of atom $j$.
Without this factor (or with $\gamma_j$$\to$0), the model
describes a sharp point-charge-density response at position
${\boldmath\mbox{$\tau$}}_j$.
With $\gamma_j$$\to$$\infty$ all local fields would be switched off,
turning the model into that of a homogeneous system again.
For a non-zero, finite value of $\gamma_j$, the factor
$\exp[i({\bf G}'-{\bf G}){{\boldmath\mbox{$\tau$}}}_j]
\exp[-\gamma_j({\bf G}'-{\bf G})^2]$ is the Fourier transform of
a Gaussian-shaped charge density 
$\sim\exp[-({\bf r}-{{\boldmath\mbox{$\tau$}}}_j)^2/(4\gamma_j)]$ 
centered at ${\boldmath\mbox{$\tau$}}_j$.
In short, this means that the charge-density response is neither
perfectly local (i.e., exactly at ${\boldmath\mbox{$\tau$}}_j$) nor 
completely delocalized (except for a truly homogeneously system, to be
characterized by $\gamma_j\to\infty$).
Instead, the charge-density response of an atom originates from its 
charge density (or at least from that of the polarizable electronic 
states), and its spatial form is included in the model.
Correspondingly, $2\sqrt{\gamma_j}$ approximates the radius of the atom.
It should be noted that the term
$\exp[-\gamma_j({\bf G}'-{\bf G})^2] \cdot 
   \exp[i({\bf G}'-{\bf G}){{\boldmath\mbox{$\tau$}}}_j]$
corresponds to the factor $\rho({\bf G}-{\bf G}')/\rho(0)$ (i.e.,
Fourier transform of the charge density) in the model by 
Bechstedt {\em et al.}\cite{Bechstedt81}
In our model the charge density of the entire system is approximated
by a composition of atomic contributions with simplified shape.

A particular role is played by the factors 
$|{\bf q}+{\bf G}||{\bf q}+{\bf G}'|$ (for metallic response) and
$({\bf q}+{\bf G})\cdot({\bf q}+{\bf G}')$ (for non-metallic response)
in Eqs. (\ref{eq5}) and (\ref{eq6}).
These factors reflect the qualitatively different origin of the
response of metallic and non-metallic systems.
For metals, long-range charge fluctuations and displacements are 
possible, resulting from intraband transitions near the Fermi level.
Such displacements lead to charge accumulation at some atoms 
and charge depletion at others.
Here our model assumes that such charge accumulation or depletion would
basically show the same spatial structure as the original charge density
of the atom (modeled by 
$\exp[-({\bf r}-{{\boldmath\mbox{$\tau$}}}_j)^2/(4\gamma_j)]$),
i.e. $\delta\rho_j({\bf r}) \sim \rho_j({\bf r})$.

The charge-density response of a non-metal, on the other hand,
is of completely different origin. 
Here the response to an external field is mainly given by a
short-range displacement of charge density from one side of the
atom to the other, i.e. by a polarization of the atom.
In many cases, this polarizability is dominated by transitions
from $s$ orbitals to $p$ orbitals or vice versa.
The spatial structure of such $s$$\leftrightarrow$$p$ polarizability is 
given by a factor
$({\bf r}$$-$${{\boldmath\mbox{$\tau$}}}_j)\cdot({\bf r}'$$-$${{\boldmath\mbox{$\tau$}}}_j)$,
leading to a factor of
$({\bf q}+{\bf G})\cdot({\bf q}+{\bf G}')$ in reciprocal space.
Again, the additional factor 
$\exp[-({\bf r}-{{\boldmath\mbox{$\tau$}}}_j)^2/(4\gamma_j)]$ 
(or its reciprocal-space counterpart) reflects the fact that the
response comes from the whole atom (including some spatial extent)
rather than from a single point.

The model can also be generalized to the case of anisotropic
response, e.g. if an atom is embedded in a non-isotropic
chemical environment, like in molecules, at surfaces, in $sp^2$-bonded 
carbon, in atomic monolayers on a substrate, or similar.
The same holds for materials with an anisotropic dielectric-constant
tensor.
In both cases, the response of each atom should be modeled with 
a direction-dependent dielectric-constant parameter
$\epsilon_j(\hat{\bf q})$.
Since such a situation can be expressed in terms of the three principal
axes ${\bf n}^{(k)}$ and corresponding principal values $\epsilon^{(k)}$ of a 
3$\times$3 tensor (see next section for details), a straight-forward 
generalization of Eq. (\ref{eq6}) is possible:
\begin{eqnarray}
\lefteqn{\mbox{atom with nonmetallic response:}} & & \nonumber\\
\lefteqn{\chi^{(j)}_{{\bf G},{\bf G}'}({\bf q}) =} & & \nonumber\\
& &   \sum_{k=1}^3
\sqrt{f(|{\bf q}+{\bf G}|;\bar{\rho}_j,\epsilon_j^{(k)})f(|{\bf q}+{\bf G}'|;\bar{\rho}_j,\epsilon_j^{(k)})} \cdot
\nonumber\\
   & & \cdot 
   [({\bf q}+{\bf G}){\bf n}_j^{(k)}]\cdot[{\bf n}_j^{(k)}({\bf q}+{\bf G}')]
      \cdot e^{-\gamma_j({\bf G}'-{\bf G})^2}
            e^{i({\bf G}'-{\bf G}){{\boldmath\mbox{$\tau$}}}_j}
\label{eq7}
\end{eqnarray}

One especially useful feature of the model proposed in this work is 
the possibility to combine metallic and non-metallic response in 
one system.
This is particularly relevant for adsorbates on metallic substrates,
for metal-insulator interfaces etc.
Here our model simply allows to attribute metallic parameters (i.e.,
Eq. (\ref{eq5})) to some atoms and non-metallic parameters
(i.e., Eqs. (\ref{eq6}) or (\ref{eq7})) to others.
For the construction of $W_{metal}$, finally, we simply take metallic
response of all atoms (i.e., Eq. (\ref{eq5})).
Apparently, for metals (or metallic regions) the dielectric function
is the same in both cases.

\subsection{Determination of the model parameters}
\label{subsec_parameters}

The determination of the parameters is a particular task.
Fortunately, the final use of the model for {\em differences} between 
metallic and non-metallic screening makes the entire approach
insensitive to the actual choice of the parameters
$V_j$, $\bar{\rho}_j$, and $\gamma_j$.
Within this work, we simply attribute a realistic volume
to each atom (for silicon, e.g., we choose $V_j$ = 20 \AA$^3$, which is
the volume per atom of bulk Si), as well as a realistic valence 
electron number (for silicon, apparently $N_j$=4).
The average electron density $\bar{\rho}_j = N_j/V_j$ defines the 
Thomas-Fermi wave number $q_{TF,j}$ and plasma frequency
$\omega_{P,j}$ for this atom, to be used in $f(Q)$.
The parameter $\gamma_j$ is obtained from least-square
fitting of the atomic charge density by a Gaussian function.
These parameters are used for both the metallic and the non-metallic
response of atom $j$.

For non-metals one needs the
dielectric-constant parameter $\epsilon_j$ (or the principal axes 
and values of the corresponding tensor for anisotropic situations).
Such values can either be taken from experiment, or they are 
calculated for the particular system.
One possibility is given by the evaluation of the small-{\bf q} limit of
$\epsilon_{{\bf G}=0,{\bf G}'=0}({\bf q})$ from the electrical-dipole 
operator applied to the interband transitions of the system, leading 
to a 3$\times$3 tensor of 
\begin{equation}
\epsilon_{ab} = \delta_{ab} + \sum_v^{occ} \sum_c^{empty} \sum_{\bf k}
   \frac{\langle v{\bf k} | \hat{p}_a | c{\bf k} \rangle 
         \langle v{\bf k} | \hat{p}_b | c{\bf k} \rangle^\ast}
             {(E_{c{\bf k}} - E_{v{\bf k}})^3}
\label{eq8}
\end{equation}
($a$,$b$ = $x$,$y$,$z$)
from which the principal axes and values can be evaluated.
Note that Eq. (\ref{eq8}) does not contain local-field effects.
However, since the resulting $\epsilon_j$ enter our model {\em before} 
the inversion of $\epsilon_{{\bf G},{\bf G}'}({\bf q})$ (which then 
leads to the local-field effects), the employment of local-field-free 
parameters is not a problem, but rather a requirement of the model.

In many systems the responses of the various atoms will differ from
each other, leading to the question of distributing the results of 
Eq. (\ref{eq8}) over the individual atoms.
Here we propose to employ an atom-centered local-orbital basis for
the calculation of the electronic states, $| n{\bf k} \rangle$.
Such a basis allows to decompose the dipole matrix elements
$\langle v{\bf k} | \hat{p}_a | c{\bf k} \rangle$ into 
individual contributions of each atom $j$, i.e. one can focus on
atom $j$ and switch off the dipole strength of all other atoms.
In this case Eq. (\ref{eq8}) yields an individual result for each
atom alone, allowing to find out the individual parameters of each
atom in the system.

\subsection{Numerical efficiency}
\label{subsec_numerics}

As an important consequence of the above
findings, the calculation of $GW$-like band structures
via $\Delta\Sigma$ allows for a tremendous gain in numerical efficiency
in four respects: ({\it i}) the use of model dielectric functions,
({\it ii}) small basis-set requirement, ({\it iii}) small band-summation
requirement, and ({\it iv}) weak influence of dynamical effects.
As mentioned, the underlying reason for all four issues is that 
the current approach 
calculates QP corrections to DFT-LDA as a {\em perturbation}.
The four issues of efficiency ({\it i})-({\it iv}) deserve detailed 
discussion.

({\it i}) The advantage of working with a dielectric model function 
rather than employing the random-phase approximation has already 
been pointed out in the last section.
In particular, the perturbative idea of Eqs. (\ref{eq_0}) and 
(\ref{eq_2})
{\em requires} a model, even beyond the issue of numerical efficiency.

\begin{figure}
\scalebox{0.45}{\includegraphics{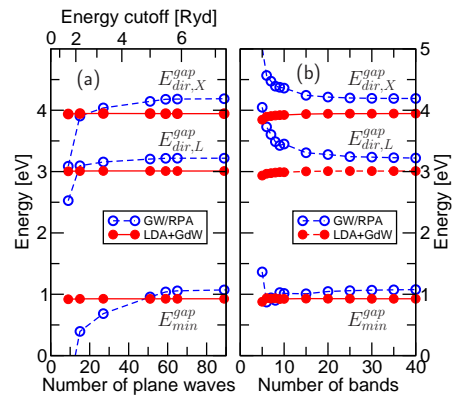}}
\caption{
\label{fig_2}
(Color online)
Gap energies of Si (indirect minimum gap, direct gap at $L$, and 
direct gap at $X$), calculated within a full $GW$ calculation 
(employing RPA screening) and within the present LDA+$GdW$ scheme.
(a) Dependence of the gap energies on the plane-wave basis 
representation of $W$ (or $(W-W_{metal})$, respectively), as controlled
by the energy cutoff (upper axis).
(b) Dependence of the gap energies on the number of bands considered
in the band summation inside the self-energy operator.
[ Note that this does not refer to the number of bands considered in the
calculation of the RPA screening.]
}
\end{figure}

({\it ii}) The basis-set requirements for 
$(W-W_{metal})$ are much weaker than for $W$ alone for two reasons:
({\it ii}.1) Within full $GW$ the bare-exchange contribution 
requires a large basis for convergence.
In our present approach, on the other hand, the bare-exchange effects 
are the same in $GW$ and $GW_{metal}$ and thus cancel each other.
({\it ii}.2) While both $W$ and $W_{metal}$ are structured in real 
space, their {\em difference} is a rather smooth function and converges
with very few basis functions.
Fig. \ref{fig_2} a shows representative gap energies of Si as 
a function of the plane-wave basis size used for $W$.
Full $GW$ requires about 
60 plane waves ($\sim$5 Ryd cutoff) for reasonable accuracy while
the LDA+$GdW$ data are already converged with
9 plane waves ($\sim$ 1.4 Ryd).
Similarly, basis-set convergence for Ar requires cutoff energies of
about 15 Ryd for GWA, but only about 2 Ryd for LDA+$GdW$.

({\it iii}) The band summation in $\Delta\Sigma$ is
less demanding than in a full $GW$ calculation because  
the influence of the higher conduction bands (via $G_1$) is weak.
This is shown in Fig. \ref{fig_2} b.
The use of about as many conduction bands as valence bands is 
sufficient for LDA+$GdW$ (at least for states near the gap), while
conventional $GW$ calculations are usually performed with at least 
about 10 times more conduction than valence bands.
This behavior again results from the smooth spatial structure
of $(W - W_{metal})$.
To summarize statements ({\it ii}) and ({\it iii}), Fig. \ref{fig_1} 
includes data from 1.4 Ryd cutoff (2.0 Ryd for Ar) and four conduction 
bands in $G_1$ as asterisks ($\ast$, ''fast'').
The agreement with the converged LDA+$GdW$ data shown by ($\bullet$)
is sufficient, except for higher-energy states.

({\it iv}) Within conventional $GW$ calculations, the correlation part 
of $\Sigma$ 
(which is subject to dynamical effects) can be as large as 5-10 eV.
Our $\Delta\Sigma$, on the other hand, is much smaller in magnitude
and thus much less sensitive to dynamical effects.
This allows to treat $\Delta\Sigma$ on the level
of the static COHSEX approximation,\cite{Hybertsen86} which we employ
in all LDA+$GdW$ calculations in this paper.
Note that this does not apply to the $GW$/RPA and $GW$/Metal reference 
calculations in this paper, all of which include dynamical effects by 
using a plasmon-pole approximation.
In those cases, the generalization of the static model 
of Sec. \ref{subsec_model} to a dynamic dielectric function is realized 
by evaluating the $f$-sum rule.\cite{Johnson74}

\section{Results for bulk silicon and argon}
\label{sec_bulk}

The QP corrections to DFT-LDA for bulk Si and Ar ar compiled in
Fig. \ref{fig_1}.
As discussed above, full $GW$ calculations using a metallic $W$ yield
QP corrections close to zero, opening the possibility of perturbative
LDA+$GdW$ as proposed in Sec. \ref{sec_method}.
In fact, the LDA+$GdW$ data are close to those of a full $GW$ calculation
employing correct, non-metallic screening from RPA (open circles).
There are, however, some deviations (related to the LDA+$GdW$ method
as such, and also to the dielectric model function).
For Si, for example, the lowest valence bands observes very small
QP corrections within $GW$/RPA, but significant negative QP corrections
within LDA+$GdW$.
Furthermore, the QP corrections for the conduction bands appear to be
less accurately reproduced by LDA+$GdW$ than for the upper valence bands.
Additional deviations are observed for the ''fast'' LDA+$GdW$ approach
(at minimal basis-set and band-summation specification), in particular
for the higher conduction bands.
These details nonwithstanding, we can conclude that LDA+$GdW$ yields
sufficient accuracy if one is interested in states near the fundamental 
gap.
Furthermore, systematic deviations between LDA+$GdW$ and $GW$/RPA can be
expected to be similar in bulk systems and other, more complicated 
systems of the same material (like, e.g., surfaces, nanostructured
systems, interfaces, etc.).
LDA+$GdW$ will allow for systematic comparison between the spectral data
of such systems.

The quite reliable LDA+$GdW$ band structures can be employed to 
yield reasonable optical spectra, as well.
To this end we include electron-hole interaction on the level
of the Bethe-Salpeter equation.\cite{Rohlfing00}
The interaction kernel is
calculated with the same (non-metallic) dielectric model 
function and same basis as the band structure.
One exception is the unscreened exchange interaction between
electrons and holes 
(originating from the Hartree potential) which
may require a larger energy cutoff than the screened interaction
and must be treated separately.
Its calculation is relatively cheap and does not affect
the efficiency of our approach.

\begin{figure}
\scalebox{0.45}{\includegraphics{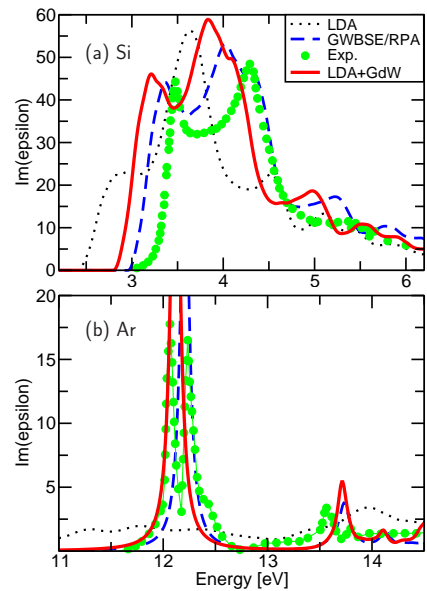}}
\caption{
\label{fig_3}
(Color online)
Optical spectrum (imaginary part of macroscopic dielectric function)
of bulk Si and of bulk Ar.
Experimental data are from 
Ref. \onlinecite{Phi72,Aspnes80,Lautenschlager87} (Si) and from 
Ref. \onlinecite{Saile76} (Ar).
Note that spin-orbit coupling (leading to the measured double-peak 
structure of the Ar exciton) is not included in our calculations.
}
\end{figure}
Fig. \ref{fig_3} shows the macroscopic imaginary dielectric function
$\epsilon_2(\omega)$ of bulk Si and Ar.
The thin dotted curves display the spectrum from the uncorrelated
LDA interband transitions, which is usually qualitatively and 
quantitatively wrong (in particular for insulators, like Ar).
The dashed lines are reference data from a full $GW$+BSE/RPA calculation,
which can be considered as the state-of-the-art approach to
$\epsilon_2(\omega)$.
The solid lines display our current results, including the
drastic numerical simplification (''fast'') as outlined above.
In comparison with the $GW$+BSE/RPA data and with experiment, the
LDA+$GdW$ data are very gratifying.
They correctly yield the two characteristic
peaks (at 3-3.5 eV and at 4-4.5 eV) of the Si spectrum. 
For Ar, we obtain an exciton peak at 12.2 eV ($GW$+BSE/RPA) or 12.1 eV
(LDA+$GdW$), respectively.
In experiment, the spin-orbit interaction (neglected in our present 
work) splits the exciton into two peaks at 12.0 and 12.2 eV.
Furthermore, a second excitonic peak is found near 13.5 eV.
The slight deviations between LDA+$GdW$ and the $GW$+BSE/RPA reference data 
mostly result from corresponding deviations in the band structure.
Note that for Ar, the agreement is even better than can
be expected from our current approach.
Most importantly, the differences between LDA+$GdW$ and $GW$+BSE/RPA are
not significantly larger than the deviations from experiment, thus 
advertising LDA+$GdW$ as a useful alternative.
Compared to the LDA interband spectrum, a tremendous improvement of
explanatory power is achieved.

\section{Results for more complex systems}
\label{sec_more}

We have tested the LDA+$GdW$ approach for a number of inhomogeneous
systems, starting from the bulk materials (Si and Ar) discussed
above.

\subsection{Si(111)-(2$\times$1) surface and silane molecule}

\begin{table}
\begin{tabular}{l@{\quad}r@{\quad}r@{\quad}r@{\quad}r@{\quad}r}
\hline
[eV]                   &  LDA    &  $GW$/  & $GW$/   & LDA+   & Exp. \\
                       &         & Metal & RPA   & $GdW$ &       \\
\hline
bulk $E^{gap}_{min}$  &  0.49  & 0.51  & 1.10  &  0.95 &  1.17$^a$ \\
$D_{up}$($J$)         &  0.0   & 0.0   &  0.0  &  0.0  &  0.0$^b$ \\
$D_{down}$($J$)       &  0.4   & 0.4   &  0.7  &  0.8  &  0.7$^c$ \\
\hline
\hline
\end{tabular}\\
$^a$Ref. \onlinecite{LB82}
$^b$Ref. \onlinecite{Uhrberg82}
$^c$Ref. \onlinecite{Perfetti87}
\caption{
\label{tab_1}
Characteristic band-structure data for the Si(111)-(2$\times$1) 
surface, which is dominated by
two dangling-bond states derived from the Pandey-chain termination.
At the $J$ point of the surface Brillouin zone the related bands
(occupied $D_{up}$ and unoccupied $D_{down}$ state) are closest to
each other and define the surface band gap.\cite{Rohlfing99}
}
\end{table}

\begin{table}
\begin{tabular}{l@{\quad}r@{\quad}r@{\quad}r@{\quad}r@{\quad}r}
\hline
[eV]                   &  LDA    &  $GW$/  & $GW$/BSE/& LDA+   & Exp. \\
              &         & Metal & RPA   & $GdW$ & [\onlinecite{Itoh86}]\\
\hline
$E_{LOMO}$            &--13.5   &--14.0 & --17.8&--17.3 &         \\
$E_{HOMO}$            & --8.4   & --9.0 & --12.5&--11.8 & --12.6 \\
$E_{LUMO}$            & --0.6   & --0.2 &    0.4&   0.6 &         \\
$\Omega_{triplet}$    &         &       &    8.0&   7.6 &         \\
$\Omega_{singlet}$    &         &       &    9.0&   8.3 &    8.8 \\
\hline
\end{tabular}
\caption{
\label{tab_2}
Spectral data of the 
SiH$_4$ molecule, which is dominated by quantum
confinement and shows the typical electronic excitations of a small
molecule.
}
\end{table}

Based on the experience with bulk silicon, we investigate two
prototypical systems of silicon in reduced dimensions, i.e. the
SiH$_4$ molecule and the Pandey-chain terminated Si(111)-(2$\times$1)
surface.
Both systems have been intensively studied in theory and 
experiment (see, e.g., Ref. \onlinecite{Rohlfing99} and 
\onlinecite{Grossman01} and references therein).
Here we focus on their electronic structure within the present
LDA+$GdW$ approach.

In the case of the Si(111)-(2$\times$1) we focus on the band
structure of the Pandey-chain derived dangling-bond 
states.\cite{Cic86,Chi71,Chi84,Uhrberg82,Perfetti87,Rei91,Nor91,Pan82}
The dangling bonds result from the lower coordination (three-fold 
instead of four-fold) of the Pandey-chain atoms,
leading to one occupied and one empty state within the bulk band
gap.\cite{Rohlfing99}
These two bands constitute one of the most intensively studied
surface electronic structures.
At the $J$ point of the surface Brillouin zone the two bands are
well separated from the silicon bulk states and define the
surface band gap.
Within LDA, this gap suffers from the same type of band-gap
underestimation as all semiconductor systems.
Here we observe a value of 0.4 eV, much smaller than the experimental
result of 0.7 eV from a combination of direct and inverse 
photoemission (see Tab. \ref{tab_1}).\cite{Uhrberg82,Perfetti87}

Within $GW$/RPA, the surface bands are significantly shifted and
result in very good agreement with the measured 
data.\cite{Uhrberg82,Perfetti87,Rei91,Nor91,Rohlfing99}
It is most gratifying to see that this behavior is also given by
the present LDA+$GdW$ approach, which yields a surface gap energy of
0.8 eV. 
This good agreement also holds for the absolute energetic position
(with respect to the bulk band structure).
Both for the occupied and for the empty band, the data from $GW$/RPA,
LDA+$GdW$, and experiment all agree to within 0.1 eV.

The screening properties for this calculation have been obtained from
the approach as outlined in Sec. \ref{subsec_parameters}, yielding
individual screening properties for each atom.
Here we find that the charge-density response of the surface atoms
is slightly larger than that of the bulk-like atoms in the center
of the slab.
The response of the bulk-like atoms agrees with that of a true bulk
calculation to within 10 percent.
At the surface, on the other hand, the smaller surface band gap, 
the $\pi$-conjugated nature of the Pandey chain, and the vicinity of 
the vacuum lead to an anisotropic response.
Perpendicular to the surface, the response is reduced by about 25 \%
(leading to a dielectric-constant parameter of about 
$\epsilon^{(\perp)}_j$=9 instead of the bulk value of $\epsilon_j$=12).
Parallel to the Pandey chain, on the other hand, the response is
doubled to $\epsilon^{(\parallel)}_j$=24.

As another, even more extreme case for silicon in reduced dimension,
we discuss the silane molecule 
(SiH$_4$).\cite{Rohlfing98,Grossman01,Itoh86}
Its electronic structure is dominated by quantum confinement,
leading to much larger band gaps and QP corrections than for
extended semiconductors.
All relevant data are compiled in Tab. \ref{tab_2}.
Compared to the LDA data, the occupied states (i.e. the lowest
occupied molecular orbital, LOMO, and (three-fold degenerate) highest
occupied molecular orbital, HOMO), are lowered in energy by more
than 4 eV. Here the current LDA+$GdW$ approach reproduces these QP shifts 
to within about 0.5 eV.
The lowest unoccupied molecular orbital (LUMO), on the other hand, 
is shifted to higher energies by 1.0 eV ($GW$/RPA) or 1.2 eV (LDA+$GdW$),
respectively.
Based on these reliable data for single-particle states,
LDA+$GdW$ also yields reasonable data for charge-neutral
electron-hole excitations (see Tab. \ref{tab_2}).
Here we take the lowest-energy singlet and triplet excitation as
representative examples.
While $GW$+BSE within RPA yields data in excellent agreement with
experiment\cite{Itoh86} and with other many-body and quantum-chemical 
methods,\cite{Grossman01}
the data from LDA+$GdW$ show slightly lower excitation energies.
The deviations are in the order of 0.5 eV and correspond to the
differences in the band-structure energy of the HOMO state, for which
LDA+$GdW$ yields a slightly too high value.
Nevertheless, in light of the huge QP corrections and very strong
electron-hole interaction of about 5 eV in SiH$_4$, we consider the 
accuracy of LDA+$GdW$ (i.e. yielding QP shifts and electron-hole binding
to within 20 \%) extremely gratifying.

Similar to the case of the Si(111)-(2$\times$1) surface, screening
in SiH$_4$ differs significantly from that of bulk silicon.
The much larger gap reduces the charge-density response strongly.
Our approach of locally evaluating the density-response contribution of 
each atom yields an isotropic response of the silicon atom to be 
described by $\epsilon_j$=3.75 (and similar results for the H atoms), 
i.e. weaker than bulk Si by a factor of 4. 
Such strong reduction for chemically passivated silicon in confined 
geometries was already found earlier.\cite{Wang94}

We close this section by mentioning that for both systems,
Si(111)-(2$\times$1) and SiH$_4$, the underlying reason for the
success of the LDA+$GdW$ approach is again given by the reproduction of
the DFT-LDA band-structure data when metallic screening is employed
in a full $GW$ calculation. 
The corresponding data are included in Tab. \ref{tab_1} and
Tab. \ref{tab_2}.
In particular for the Si(111)-(2$\times$1) surface, this
mandatory condition for the applicability of LDA+$GdW$ is 
nearly exactly fulfilled.
For the SiH$_4$ molecule some difference of the order of 0.5 eV
are found.
Considering the massive deviation of this system from a homogeneous
metal, even this agreement to within 0.5 eV is an amazing result.

\subsection{Argon systems}

\begin{figure}
\scalebox{0.45}{\includegraphics{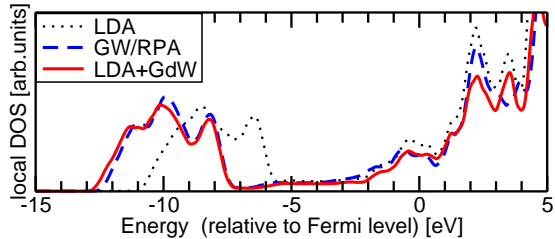}}
\caption{
\label{fig_4}
(Color online)
Local density of states in a monolayer of Ar,
combined with a 5-layer aluminium(001) slab in a heterostructure.
}
\end{figure}
Spatially varying dielectric response is also present in 
metal-insulator heterostructures.
As an example 
Fig. \ref{fig_4} shows the single-particle spectrum
of a periodic heterostructure composed of five atomic layers (10 \AA) 
of aluminium and one atomic layer (3 \AA) of argon, stacked along the
Al(001) direction.
This system combines metallic screening in Al with insulating 
behavior in Ar, which has significant consequences on the QP 
energetics.\cite{Charlesworth93,Wang04,Neaton06,Thygesen09} 
Here we focus on the local density of states (LDOS) in the Ar
monolayer.
The LDOS between --12 eV and --6 eV results from the upper valence
states of Ar (3$p$), while the LDOS above 2 eV comes from
the Ar conduction bands, with increasing admixture of Al states at
higher energy.
The LDOS inside the Ar band gap (--6 eV to +2 eV) results from
spill-out of Al states into the Ar layer.
The most interesting feature is the rather small QP correction of the
argon states, which (in $GW$/RPA) amounts to --1.7 eV (+0.2 eV) for
the upper valence (lower conduction) states, yielding a total
correction of 1.9 eV for the fundamental gap of Ar.
In bulk Ar, on the other hand, the gap-edge states observe QP shifts
of --4.1 eV and +2.0 eV, yielding a gap correction of
6.1 eV (cf. Fig. \ref{fig_1}).
The presence of metallic screening in the immediate neighborhood
significantly weakens the QP shifts due to image-state 
effects,\cite{Charlesworth93,Wang04,Neaton06,Thygesen09} 
both for holes and for electrons (by about 2 eV each).
It is most gratifying to see that in our LDA+$GdW$ approach
(again with a plane-wave cutoff of 2 Ryd), the spatial set-up of the 
dielectric model function (cf. Eq. (\ref{eq_3})) reproduces these 
effects. 
Here LDA+$GdW$ yields QP shifts of --1.8 eV and
+0.2 eV for the band-edge states and a gap correction of +2.0 eV,
compared to the Ar bulk data (see Fig. \ref{fig_1}) of --3.5 eV,
+2.8 eV, and +6.3 eV.
We conclude that the LDA+$GdW$ approach is a suitable method for 
addressing electronic properties of metal-nonmetal junctions.

\begin{figure}
\scalebox{0.45}{\includegraphics{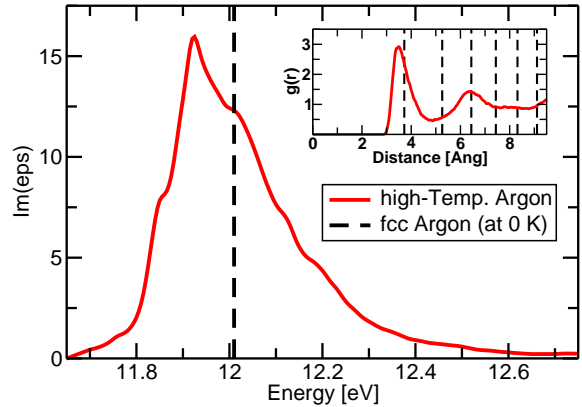}}
\caption{
\label{fig_5}
(Color online)
Optical spectrum of the excitons in non-crystalline argon (from
molecular dynamics of 64 atoms at 300 Kelvin, at the solid-argon 
density --- see text), resulting from the current LDA+$GdW$ approach.
The dashed line indicates the position of the exciton in the periodic
crystal.
The inset shows the Ar-Ar pair-correlation function of the MD 
simulation.
}
\end{figure}
As a last example for the potential of our method, Fig. \ref{fig_5} 
shows the exciton spectrum of non-crystalline argon.
At zero temperature argon forms a periodic face-centered cubic (fcc) 
lattice, which can easily be treated by MBPT (see 
Sec. \ref{sec_method}), leading to the results as discussed in 
Sec. \ref{sec_bulk}.
For this periodic solid the exciton yields a sharp line (except for
dynamical broadening effects from self trapping or similar, that are 
completely neglected here).
This changes in the case of non-periodic argon, like
in its liquid or amorphous state.
Such systems may be described by sufficiently large supercells.
At present we investigate the spectra resulting from a 64-atom cell
(consisting of 4$\times$4$\times$4 fcc unit cells) and exploit its
spectral features from the $\Gamma$ point of the supercell, only.
For the periodic fcc crystal this yields an exciton at 12.01 eV
excitation energy (slightly lower than the value reported in Sec.
\ref{sec_bulk}, which was obtained from the standard fcc unit 
cell containing one atom, and 500 {\bf k}-points in the BSE).
Within this configuration (which is computationally much more demanding
than a simple one-atom-fcc calculation and is extremely demanding
for the standard $GW$+BSE Hamiltonian) the spectrum of liquids or
amorphous systems can be evaluated.
At present we simply consider argon at its solid-state density
(for comparison sake), but 
heated to 300 Kelvin (although this is an unrealistic
high temperature for argon at this density).
We perform a constant-temperature molecular-dynamics simulation (using 
a simple Lennard-Jones interatomic potential), leading to the
Ar-Ar pair-correlation function shown in the inset of 
Fig. \ref{fig_5}.
Such a simulation is certainly not fully realistic in terms of 
describing liquid or amorphous systems; nonetheless it yields 
structural elements that may very well be present in liquids.
The pair-correlation function clearly exhibits structures beyond
harmonic vibrations 
(like, e.g. the vanishing of the second-nearest-neighbor peak 
at 5.3 \AA), thus prohibiting a perturbative electron-phonon 
interaction treatment in the evaluation of the spectrum.
Instead, our LDA+$GdW$ approach (averaged over 20 snapshots of the 
MD run) easily allows to evaluate the spectrum.
The data shown in Fig. \ref{fig_5} clearly demonstrate three 
important features:
(i) the exciton line is significantly broadened,
(ii) the broadening is asymmetric, leading to substantial
non-zero amplitude well above the exciton energy, and
(iii) the maximum of the peak is at lower energy than in the
periodic system.
The third feature is related to the fact that in the
pair-correlation function, the first maximum  also occurs at smaller 
distance (3.4 \AA) than the fcc nearest-neighbor-distance (3.7 \AA),
which is a consequence of the anharmonicity of the Ar-Ar interatomic 
potential.

\section{Summary}
\label{sec_summary}

In summary, we have discussed an extremely efficient modification
of standard many-body perturbation theory ($GW$ method plus 
Bethe-Salpeter equation).
Based on the observation that metallic screening in the $GW$ method 
approximately reproduces the DFT-LDA band structure (which should be 
checked for 
each material), quasiparticle (QP) corrections to DFT-LDA are obtained 
in a truly perturbative approach at minimal cost, provided that the
dielectric screening can be described by an appropriate model.
The resulting QP band structures and optical spectra (including 
electron-hole interaction) are slightly less accurate than those
from conventional $GW$+BSE, but they include all Coulomb-interaction
effects (like screening, electron-hole binding etc.) in a physically
correct way, allowing to systematically investigate excitations 
beyond DFT and beyond the computational limits of conventional MBPT.

\end{document}